\documentclass[12pt,preprint,eqsecnum]{aastex}

\usepackage{emulateapj5}

\usepackage{graphicx}
\usepackage{dcolumn}
\usepackage{bm}
\usepackage{epsf}

\def\Lie{\mathcal{L}}

\shortauthors{BAUMGARTE \& SHAPIRO}
\shorttitle{GENERAL RELATIVISTIC MHD}

\begin{document}

\title{General-Relativistic MHD for the Numerical Construction of 
	Dynamical Spacetimes}

\author{Thomas W. Baumgarte \altaffilmark{1,2} and 
	Stuart L. Shapiro \altaffilmark{2,3}}

\affil{\altaffilmark{1}
	Department of Physics and Astronomy, Bowdoin College,
	Brunswick, ME 04011}

\affil{\altaffilmark{2}
	Department of Physics, University of Illinois at
	Urbana-Champaign, Urbana, Il~61801}

\affil{\altaffilmark{3}
	Department of Astronomy and NCSA, University of Illinois at
     	Urbana-Champaign, Urbana, Il~61801}

\begin{abstract}
We assemble the equations of general relativistic magnetohydrodynamics
(MHD) in $3+1$ form. These consist of the complete coupled set of Maxwell
equations for the electromagnetic field, Einstein's equations for the
gravitational field, and the equations of relativistic MHD for a
perfectly conducting ideal gas. The adopted form of the equations is
suitable for evolving numerically a relativistic MHD fluid in a 
dynamical spacetime characterized by a strong gravitational field.
\end{abstract}

\maketitle

\section{Introduction}
\label{sec1}

Magnetic fields play a crucial role in determining the evolution of
many relativistic objects.  In any highly conducting astrophysical
plasma, a frozen-in magnetic field can be amplified appreciably by gas
compression or shear.  Even when an initial seed field is weak,
the field can grow in the course of time to significantly influence the gas
dynamical behavior of the system. If, in addition, the gravitational
field is strong and dynamical, magnetic fields can even affect the
entire geometry of spacetime, according to general relativity. 
In this situation, terms involving magnetic and
electric fields are important not only as electromagnetic forces
acting on the matter in the equations of relativistic hydrodynamics
but also as stress-energy sources governing the metric in Einstein's
gravitational field equations.

In this paper (hereafter Paper I) we assemble the complete set of
Maxwell--Einstein--magnetohydrodynamic equations that determine the
self-consistent evolution of a relativistic, ideal MHD fluid in a
dynamical spacetime. Our goal is to set down a formulation of the
equations that is suitable for numerical integration in full $3+1$
dimensions.  Subsets of these equations have appeared elsewhere, but
we recompile the complete set here for convenience and future
reference.  We reconcile several seemingly different, but equivalent,
forms of the equations that have appeared in the existing literature.
We also correct some errors in previously published results.  In a
companion paper (Baumgarte \& Shapiro, 2002, hereafter Paper II) we
use these general relativistic MHD equations to follow the
gravitational collapse of a magnetized star to a black hole.

We are motivated in part by the growing list of important, unsolved
problems which involve hydromagnetic effects in strong-field dynamical
spacetimes.  The final fate of many of these astrophysical systems,
and their distinguishing observational signatures, hinge on the role
that magnetic fields may play during the evolution. Some of these
systems are promising sources of gravitational radiation for detection
by laser interferometers now under design and construction, like LIGO,
VIRGO, TAMA, GEO and LISA. Others may be responsible for gamma-ray
bursts. Recent examples of astrophysical scenarios involving
strong-field dynamical spacetimes in which MHD effects may play a
decisive role include the following:

$\bullet$ The merger of binary neutron stars. The merger can lead to
the formation of a {\it hypermassive} star supported by differential
rotation (Baumgarte, Shapiro \& Shibata 2000; Shibata \& Uryu 2000).
While such a star may be dynamically stable against gravitational
collapse and bar formation, the radial stabilization due to
differential rotation is likely to be temporary. Magnetic braking and
viscosity, driven by differential rotation, combine to drive the star
to uniform rotation, even if the seed magnetic field and the viscosity
are small (Shapiro 2000). This process inevitably leads to delayed
collapse, which will be accompanied by a delayed gravitational wave
burst.

$\bullet$ Core collapse in a supernova. Core collapse may again induce
differential rotation, even if the rotation of the progenitor at the
onset of collapse is only moderately rapid and almost uniform (see,
e.g., Zwerger \& M\"{u}ller 1997; Rampp, M\"{u}ller \& Ruffert 1998, and
references therein).  Differential rotation can wind up a frozen-in
magnetic field to high values, at which point it may provide a
significant source of stress. Hypermassive neutron stars may form and
survive until the fields are weakened by magnetic braking or other
instabilities.

$\bullet$ The generation of gamma-ray bursts (GRBs). Typical 
models for GRB formation
involve the collapse of rotating massive stars to a black hole
(MacFadyen \& Woosley 1999) or the merger of binary neutron stars
(Narayan, Paczynski \& Piran 1992) or the tidal disruption of a
neutron star by a black hole (Ruffert \& Janka 1999).  In current
scenarios, the burst is powered by the extraction of rotational energy
from the neutron star or black hole, or from the remnant disk material
formed about the black hole (Vlahkis \& Konigl 2001). Strong magnetic
fields provide the likely mechanism for extracting this energy on the
required timescale and driving collimated GRB outflows in the form of
relativistic jets (Meszaros \& Rees 1997; Sari, Piran \& Halpern 1999;
Piran 2002).  Even if the initial magnetic fields are weak, they can
be amplified to the required values by differential motions or dynamo
action.

$\bullet$ Supermassive star (SMS) collapse. SMSs may form in the early
universe and their catastrophic collapse may provide the origin of
supermassive black holes (SMBHs) observed in galaxies and quasars (see Rees
1984 and Baumgarte \& Shapiro 1999a for discussion and references) If
a SMS is uniformly rotating, cooling and secular 
contraction will ultimately trigger its
coherent dynamical collapse to a SMBH, giving rise to a burst of
gravitational waves (Saijo et al. 2002; Shibata \& Shapiro 2002).  If a 
SMS is differentially rotating, cooling and contraction will instead 
lead to the unstable formation of bars or spiral arms 
prior to collapse, and the production of quasi-periodic waves
(New \& Shapiro 2001a,b). Magnetic fields and turbulence provide the
principle mechanisms that can damp differential rotation in such stars
(Zel'dovich \& Novikov 1971; Shapiro 2000) and thus determine 
their ultimate fate.

$\bullet$ The r-mode instability in rotating neutron stars. This
instability has recently been proposed as a possible mechanism for
limiting the angular velocities in neutron stars and producing
observable quasi-periodic gravitational waves (Andersson 1998;
Friedman \& Morsink 1998; Andersson, Kokkotas \& Stergioulas 1999;
Lindblom, Owen \& Morsink 1998).  However, preliminary calculations
(Rezzolla et al. 2000, 2001a,b and references therein) suggest that if
the stellar magnetic field is strong enough, r-mode oscillations will
not occur. Even if the initial field is weak, fluid motions produced
by these oscillations may amplify the magnetic field and eventually
distort or suppress the r-modes altogether. [R-modes may also by
suppressed by non-linear mode coupling (Arras et al. 2002; Schenk et
al. 2002).]

This paper is partitioned as follows: In Sections \ref{sec2} and
\ref{sec3} we review Einstein's field equations and Maxwell's
equations in $3+1$ form.  In Section \ref{sec4} we discuss the
approximation of ideal magnetohydrodynamics and in Section
\ref{sec5} the equations of general relativistic hydrodynamics.  
In Section \ref{sec6} we then develop the equations of general 
relativistic MHD.  We derive the MHD source terms that appear in 
Einstein equations in Section \ref{sec7}. We compare our results with 
those of previous treatments in Section \ref{sec8}.  Finally, we 
briefly summarize our analysis in Section \ref{sec9}.
 
We adopt geometrized units throughout, setting $G=1=c$, where $G$ is
the gravitation constant and $c$ is the speed of light.

\section{Einstein's Field Equations in 3+1 Form}
\label{sec2}

The spacetime geometry (i.e.~metric) is determined by integrating
Einstein's field equations of general relativity. Most algorithms for
performing this integration numerically are based on a $3+1$
decomposition of Einstein's equations, which is ideally suited for
solving the general initial value problem. Below we briefly summarize
the key equations that result from recasting Einstein's equations in
$3+1$ form. More detailed discussions may be found, for example, in
Misner, Thorne \& Wheeler (1973), York (1979), Evans (1984) and
references therein.

In a $3+1$ decomposition of Einstein's field equations, the
four-dimensional spacetime $M$ is foliated into a family of
non-intersecting spacelike three-surfaces $\Sigma$, which arise, at
least locally, as level surfaces of a scalar time function $t$.  The
spatial metric $\gamma_{ab}$ on the three-dimensional hypersurfaces
$\Sigma$ is induced by the spacetime metric $g_{ab}$ according to
\begin{equation} \label{projection}
\gamma_{ab} = g_{ab} + n_a n_b,
\end{equation}
where $n^a$ is the unit normal vector $n_a = \alpha \nabla_a t$ to the
slices.  Here the normalization factor $\alpha$ is called the lapse
function.  The time vector $t^a$ is constructed so that it is dual to
the foliation 1-forms $\nabla_a t$:
\begin{equation} \label{t}
t^a = \alpha n^a + \beta^a,
\end{equation}
where the shift vector $\beta^a$ is spatial, i.e., $n_a \beta^a = 0$, but
otherwise arbitrary.  In a coordinate system that is aligned with $t^a$
and $\Sigma$, the components of $n^a$ are
\begin{equation} \label{n}
n_a = (-\alpha,0,0,0) \mbox{~~~and~~~} n^a = \alpha^{-1}(1,-\beta^i).
\end{equation}
We adopt the convention that roman indices $a, b, c, d, \ldots$ denote
spacetime components, while $i, j, k, l, \ldots$ denote spatial
components.  

The spacetime metric can now be written in the ADM form (Arnowitt,
Deser \& Misner 1962)
\begin{equation}
ds^2 = - \alpha^2 dt^2 + \gamma_{ij} (dx^i + \beta^i dt)(dx^j +
\beta^j dt).
\end{equation}
The lapse function $\alpha$ determines by how much proper time
advances along the normal vector from one time slice to the next, and
the shift vector $\beta^i$ determines by how much spatial coordinates
are shifted on the new slice.  The lapse function and three components of 
the shift vector constitute gauge potentials that may be
freely specified. Together, $\alpha$ and $\beta^i$ thus embody the
four degrees of coordinate freedom inherent of general relativity.

Einstein's equations
\begin{equation}
G_{ab} = 8 \pi T_{ab},
\end{equation}
where $G_{ab}$ is the Einstein tensor associated with $g_{ab}$ and
$T_{ab}$ is the stress-energy tensor, can be projected both along the
normal direction $n^a$ and into the spatial slice $\Sigma$.  The
spatial projection yields two constraint equations, which constrain
the fields within each slice $\Sigma$; 
they contain at most one time derivative of the
spatial metric.  The projection along the normal vector yields an
evolution equation that describes how the fields propagate from one
slice to the next; it contains second-order time derivatives of the
spatial metric.  The constraint equations consist of the Hamiltonian
constraint
\begin{equation} \label{ham}
R + K^2 - K_{ij} K^{ij} = 16 \pi \rho
\end{equation}
and the momentum constraint
\begin{equation} \label{mom}
D_j (K^{ij} - \gamma^{ij} K ) = 8 \pi S^i,
\end{equation}
and the evolution equation is
\begin{equation}  \label{kdot}
\begin{array}{rcl}
\partial_t K_{ij} & = & - D_i D_j \alpha + \alpha (R_{ij} - 2 K_{ik} K^k_{~j}
+ K K_{ij}) \\[1mm]
& & - 8 \pi \alpha(S_{ij} - \frac{1}{2} \gamma_{ij} (S - \rho)) 
+ \Lie_{\bf \beta} K_{ij}.
\end{array}
\end{equation}
Here $K_{ij}$ is the extrinsic curvature, and its definition in terms
of the time derivative of the spatial metric is usually considered the
second evolution equation
\begin{equation} \label{gdot}
\partial_t \gamma_{ij} = -2 \alpha K_{ij} + \Lie_{\bf \beta} \gamma_{ij}.
\end{equation}
Here $D_i$, $R_{ij}$ and $R = \gamma^{ij} R_{ij}$ are the covariant
derivative, Ricci tensor and scalar curvature associated with
$\gamma_{ij}$, while $K = \gamma^{ij} K_{ij}$ is  
the trace of the extrinsic curvature. 
The symbol $\Lie$ denotes a Lie derivative. 
The matter and nongravitational field sources $\rho$, $S_i$ and $S_{ij}$
are the projections of the stress-energy tensor into $n^a$ and $\Sigma$
and are given by
\begin{eqnarray}
\rho & = & n_a n_b T^{ab} \label{rho} \\
S_i & = & - \gamma_{ia} n_b T^{ab} \label{Si} \\
S_{ij} & = & \gamma_{ia} \gamma_{jb} T^{ab}. \label{Sij}
\end{eqnarray}
The quantity $\rho$ is the total mass-energy density as measured by a normal
observer, $S_i$ is the momentum density and $S_{ij}$ is the stress.
Finally, $S$ is defined as the trace of $S_{ij}$, 
\begin{equation} \label{trS}
S = \gamma^{ij} S_{ij}. 
\end{equation}
We remark that if the constraint equations are satisfied on an initial
time slice $\Sigma$, the evolution equations guarantee that the 
constraints will be satisfied on all subsequent time slices.

Equations (\ref{ham}) through (\ref{gdot})
are commonly referred to as the ADM equations (Arnowitt, Deser \&
Misner, 1962).  Numerical implementations of these equations have
revealed that their numerical stability can be improved dramatically
by bringing them into a slightly different form.  One such
modification, now commonly referred to as ``BSSN'', is based on
Shibata \& Nakamura (1995) and Baumgarte \& Shapiro (1999b).  This
system is widely used, and its enhanced stability properties have been
analyzed by several authors (e.g.~Alcubierre et al., 2000; see Knapp,
Walker \& Baumgarte 2002 for an electromagnetic analogy).
Alternatively, several authors have experimented with hyperbolic
formulations of Einstein's equations (e.g. Anderson \& York 1999).
We refer the reader to these papers for further details and references.

\section{Maxwell's Equations}
\label{sec3}

We decompose the Faraday tensor $F^{ab}$ as
\begin{equation} \label{faraday1}
F^{ab} = n^a E^b - n^b E^a + \epsilon^{abc} B_c
\end{equation}
so that $E^a$ and $B^a$ are the electric and magnetic fields observed
by a normal observer $n^a$.  Both fields are purely spatial, whereby
\begin{equation}
E^a n_a = 0 \mbox{~~~~and~~~~} B^a n_a = 0,
\end{equation}
and the three-dimensional Levi-Civita symbol $\epsilon_{abc}$ is
defined by
\begin{equation} \label{epsilon}
\epsilon^{abc} = \epsilon^{abcd} n_d 
\mbox{~~~or~~~}
\epsilon_{abc} = n^d \epsilon_{dabc}.
\end{equation}
Note that $\epsilon^{abc}$ is non-zero only for spatial indices, while
$\epsilon_{abc}$ may be non-vanishing even if one index is timelike
(see equation (\ref{epsilon_convert}) below).  We also decompose the
electromagnetic current four-vector $\mathcal{J}^a$ according to
\begin{equation} \label{defJ}
\mathcal{J}^a = n^a \rho_e + J^a,
\end{equation}
where $\rho_e$ and $J^a$ are the charge density and current as
observed by a normal observer $n^a$. Note that $J^a$ is purely
spatial, $J^a n_a = 0$.

With these definitions, Maxwell's equations,
\begin{equation} \label{maxwell1}
\nabla_b F^{ab} = 4 \pi \mathcal{J}^a
\end{equation}
and
\begin{equation} \label{maxwell2}
\nabla_{[a} F_{bc]} = 0,
\end{equation}
where $\nabla$ is the four-dimensional covariant derivative operator
associated with $g_{ab}$, can be brought into the $3+1$ form
\begin{eqnarray}
D_i E^i & = & 4 \pi \rho_e \label{divE} \\
\partial_t E^i & = & \epsilon^{ijk} D_j ( \alpha B_k)
	- 4 \pi \alpha J^i + \alpha K E^i + \Lie_{\bf \beta} E^i
	\label{Edot} \\
D_i B^i & = & 0 \label{divB} \\
\partial_t B^i & = & - \epsilon^{ijk} D_j (\alpha E_k) + \alpha K B^i
	+ \Lie_{\bf \beta} B^i  \label{Bdot}
\end{eqnarray}
(see, e.g., Thorne \& MacDonald 1982).  The charge conservation
equation,
\begin{equation}
\nabla_a \mathcal{J}^a = 0,
\end{equation}
which is implied by (\ref{maxwell1}), becomes
\begin{equation}
\partial_t \rho_e = - D_i (\alpha J^i) + \alpha K \rho_e 
	+ \Lie_{\bf \beta} \rho_e. \label{rhodot}
\end{equation}
The special relativistic Maxwell's equations can be recovered very
easily by evaluating the above equations for a Minkowski spacetime
with $\gamma_{ij} = f_{ij}$, where $f_{ij}$ is the flat spatial metric
in an arbitrary coordinate system, $\alpha = 1$, $K = 0$ and $\beta^i
= 0$.

It is convenient to introduce a four-vector potential $\mathcal{A}_a$, which
can be decomposed into
\begin{equation}
\mathcal{A}_a = \Phi n_a + A_a,
\end{equation}
where $A_a$ is purely spatial, $A_a n^a = 0$.  Inserting (\ref{n}),
this implies
\begin{equation} \label{A_t}
A_t = \beta^i A_i
\end{equation}
while $A^t = 0$, as for any spatial vector.  In terms of the vector
potential, the Faraday tensor can be written
\begin{equation} \label{faraday2}
F_{ab} = \mathcal{A}_{b,a} - \mathcal{A}_{a,b} = n_a E_b - n_b E_a +
	\epsilon_{abc} B^c
\end{equation}
Contracting this equation with $\epsilon^{abc}$ yields
\begin{equation}
\epsilon^{abc} (\mathcal{A}_{b,a} - \mathcal{A}_{a,b}) =
\epsilon^{abc} \epsilon_{abd} B^d = 2 B^c
\end{equation}
or
\begin{equation} \label{B}
B^i = \epsilon^{ijk} A_{k,j}.
\end{equation}
Note that with this identification, the magnetic field $B^i$ 
automatically satisfies the constraint equation (\ref{divB}).

It is possible, and often convenient (see Paper II), to re-write
Maxwell's equations completely in terms of $E_i$ and $A_i$, thereby
eliminating $B_i$.  Evaluating (\ref{faraday2}) for the components $a
= t$ and $b = i$ with $\mathcal{A}_i = A_i$ and $\mathcal{A}_t = -
\alpha \Phi + \beta^i A_i$ yields
\begin{equation}
\partial_t A_i = - \alpha E_i + \epsilon_{tij} B^j 
	- (\alpha \Phi - \beta^j A_j),_i.
\end{equation}
Using the definition (\ref{epsilon}), we can rewrite
\begin{eqnarray} \label{epsilon_convert}
\epsilon_{tij} & = & n^d \epsilon_{dtij} 
	= - \alpha^{-1} \beta^k \epsilon_{ktij} 
	= - \alpha^{-1} \beta^k \epsilon_{tikj} \nonumber \\
	& = & - \beta^k n^d \epsilon_{dikj} = -\beta^k \epsilon_{ikj},
\end{eqnarray}
so that 
\begin{equation} \label{Adot2}
\partial_t A_i = - \alpha E_i + \epsilon_{ijk} \beta^j B^k 
	- (\alpha \Phi - \beta^j A_j),_i.
\end{equation}
With (\ref{B}), $\epsilon_{ijk} \beta^j B^k$ can be expressed in terms
of $A_i$ as
\begin{eqnarray}
\epsilon_{ijk} \beta^j B^k 
& = & \epsilon_{ijk} \epsilon^{klm} \beta^j A_{m,l} \nonumber \\
& = & (\delta^l_{~i} \delta^m_{~j} - \delta^l_{~j} \delta^m_{~i}) 
	\beta^j A_{m,l} \nonumber \\
& = & \beta^j A_{j,i} - \beta^j A_{i,j}.
\end{eqnarray}
Inserting this into (\ref{Adot2}) yields
\begin{equation} \label{Adot}
\partial_t A_i = - \alpha E_i - \partial_i (\alpha \Phi)
+ \Lie_{\bf \beta} A_i.
\end{equation}
In equation (\ref{Edot}), the magnetic field $B_i$ can be eliminated
similarly:
\begin{eqnarray}
\epsilon^{ijk} D_j ( \alpha B_k) & = &
\epsilon^{ijk} D_j ( \alpha \epsilon_{klm} D^l A^m ) \nonumber \\
& = & \epsilon^{ijk}  \epsilon_{klm} D_j (\alpha D^l A^m) \nonumber \\
& = & (\delta^i_{~l} \delta^j_{~m} - \delta^i_{~m} \delta^j_{~l})
	 D_j (\alpha D^l A^m) \nonumber \\
& = & D_j (\alpha D^i A^j) - D_j (\alpha D^j A^i).
\end{eqnarray}
Inserting this into (\ref{Edot}) yields
\begin{equation} \label{Edot2}
\partial_t E^i = D_j (\alpha D^i A^j) - D_j (\alpha D^j A^i) 
  - 4 \pi \alpha J^i + \alpha K E^i + \Lie_{\bf \beta} E^i
\end{equation}
Equations (\ref{Adot}) and (\ref{Edot2}) form a system of equations
for $E^i$ and $A_i$ alone.  In the special relativistic limit, they
again reduce to familiar expressions.  We also note that in terms of
partial derivatives, equation (\ref{Edot2}) can be expanded to yield
\begin{eqnarray} \label{Edot3}
\partial_t E^i & = & \gamma^{-1/2} \left( \alpha \gamma^{1/2} (\gamma^{il}
\gamma^{jm} - \gamma^{im} \gamma^{jl}) A_{m,l} \right),_j \nonumber \\
& & - 4 \pi \alpha J^i + \alpha K E^i + \Lie_{\bf \beta} E^i,
\end{eqnarray}
where $\gamma$ is the determinant of the spatial metric $\gamma_{ij}$.
This form of the electric field evolution equation will be useful for
applications in Paper II.

\section{Ideal Magnetohydrodynamics Approximation}
\label{sec4}

Ohm's law can be written (see, e.g., Jackson, 1999)
\begin{equation} \label{ohm}
{\mathcal J}_a - \tilde \rho_e u_a = \sigma F_{ab} u^b,
\end{equation}
where $\sigma$ is the electrical conductivity and $\tilde \rho_e =
-{\mathcal J}^a u_a$ is the charge density as seen by an observer
co-moving with the fluid four-velocity $u^a$ (in contrast to $\rho_e$,
which was defined as the charge density as observed by a normal
observer $n^a$).

A $3+1$ decomposition of Ohm's law can be derived by contracting
(\ref{ohm}) with $n^a$ and $\gamma_a^{~b}$.  The former yields
\begin{equation} \label{ohm_time}
W \tilde \rho_e = \rho_e - \sigma u_a E^a,
\end{equation}
where we have defined $W$ as the Lorentz factor between normal and
fluid observers
\begin{equation} \label{W}
W \equiv - n_a u^a = \alpha u^t.
\end{equation}
Projecting (\ref{ohm}) into $\Sigma$, or, equivalently, evaluating
the spatial components $a = i$ of (\ref{ohm}), yields
\begin{equation}
\begin{array}{rcl} \label{ohm_spatial}
J_i - \tilde \rho_e u_i & = & 
\sigma F_{ia} u^a = \sigma (F_{it} u^t + F_{ij} u^j) \\
& = & \sigma(\alpha E_i u^t + \epsilon_{itk} B^k u^t 
	+ \epsilon_{ijk} B^k u^j)\\
& = & \sigma \left( W E_i + \epsilon_{ijk} (\beta^j B^k u^t + B^k u^j)\right)\\
& = & \sigma \left( W E_i + \epsilon_{ijk} (v^j + \beta^j) B^k u^t) \right).
\end{array}
\end{equation}
Here we have defined
\begin{equation} \label{v}
v^i \equiv \frac{u^i}{u^t}
\end{equation}
and have used (\ref{epsilon_convert}) to relate $\epsilon_{itk}$ to
$\epsilon_{ijk}$, giving rise to the shift term in (\ref{ohm_spatial})
(the shift term is missing in some previous treatments, see Section
\ref{sec8}).

Dividing Ohm's law (\ref{ohm}) by $\sigma$ and allowing $\sigma
\rightarrow \infty$ yields the perfect conductivity condition
\begin{equation} \label{ideal_MHD}
F_{ab} u^b = 0.
\end{equation}
According to (\ref{ohm_time}) and (\ref{ohm_spatial}), this result is
equivalent to the condition that the electric field vanish in 
the fluid rest frame
\begin{equation}
u_a E^a = 0,
\end{equation}
or
\begin{equation} \label{MHD}
\alpha E_i = - \epsilon_{ijk} (v^j + \beta^j) B^k,
\end{equation}
which is often called the ideal MHD relation.  When evaluated in a
Minkowski spacetime, the last equation reduces to the familiar
expression $E_i = - \epsilon_{ijk} v^j B^k$ or 
${\bf E = -v \times B}$.

We can now evaluate Faraday's law (\ref{Bdot}) under the assumption of
perfect conductivity.  Taking the trace of (\ref{gdot}) yields
\begin{equation}
\alpha K = - \partial_t \ln \gamma^{1/2} + D_i \beta^i.
\end{equation}
The above expression can be combined with the Lie derivative 
$\Lie_{\bf \beta} B^i$ to give
\begin{equation} \label{rhs}
\alpha K B^i + \Lie_{\bf \beta} B^i = D_j (\beta^j B^i - \beta^i B^j)
	- B^i \partial_t \ln \gamma^{1/2},
\end{equation}
where we have used (\ref{divB}).  Inserting (\ref{rhs}) together with
the ideal MHD equation (\ref{MHD}) into Faraday's law (\ref{Bdot})
reveals that all the shift terms cancel, leaving
\begin{equation}  \label{MHD_faraday}
\frac{1}{\gamma^{1/2}} \partial_t (\gamma^{1/2} B^i)
= D_j (v^i B^j - v^j B^i).
\end{equation}

It is convenient to introduce the magnetic vector density 
\begin{equation}
{\mathcal B}^i \equiv \gamma^{1/2} B^i,
\end{equation}
in terms of which equation (\ref{MHD_faraday}) and (\ref{divB}) 
reduce to the particularly simple forms
\begin{equation} \label{MHD_faraday2}
\partial_t {\mathcal B}^i = 
	\partial_j (v^i {\mathcal B}^j - v^j {\mathcal B}^i),
\end{equation}
and
\begin{equation}
\partial_i {\mathcal B}^i = 0.
\end{equation}

In Section \ref{sec8} we compare with previous treatments and 
correct errors in some previously published equations.

\section{General Relativistic Hydrodynamics}
\label{sec5}

For a perfect fluid, the stress-energy tensor $T^{ab}_{\rm fluid}$ can
be written
\begin{equation} \label{T_fluid}
T^{ab}_{\rm fluid} = \rho_0 h u^a u^b + P g^{ab}.
\end{equation}
Here $\rho_0$ is the rest-mass density as observed by an observer
co-moving with the fluid $u^a$, $P$ is the pressure, and $h$ is the
specific enthalpy
\begin{equation}
h = 1 + \epsilon + P/\rho_0,
\end{equation}
where $\epsilon$ is the specific internal energy density.

In the absence of electromagnetic fields, the equations of motion for the 
fluid can be derived
from the local conservation of energy-momentum,
\begin{equation} \label{divT}
\nabla_b T^{ab}_{\rm fluid} = 0,
\end{equation}
and the conservation of baryons,
\begin{equation} \label{cont}
\nabla_a (\rho_0 u^a) = 0.
\end{equation}
The resulting equations can be cast in various forms, depending on how
the the primitive fluid variables are chosen (see, e.g., Font 2000 for
a recent review).  The most frequently adopted relativistic formalism
was originally developed by Wilson (1972; see also Hawley, Smarr \&
Wilson, 1984), who defined a rest-mass density variable
\begin{equation}
D \equiv \rho_0 W,
\end{equation}
an internal energy density variable
\begin{equation}
E \equiv \rho_0 \epsilon W,
\end{equation}
and a momentum variable
\begin{equation}
S_a \equiv \rho_0 h W u_a = (D + E + PW) u_a.
\end{equation}
Note that the spatial vector $S_i$ defined above is the fluid
contribution to the source term $S_i$ appearing in Einstein's field
equations (cf., equation (\ref{Si})).  In terms of these variables,
the equation of continuity becomes
\begin{equation}
\partial_t (\gamma^{1/2} D) 
+ \partial_j (\gamma^{1/2} D v^j) 
=  0. \label{DWdot} 
\end{equation}
Contracting (\ref{divT}) with $u^b$ yields the energy equation
\begin{eqnarray}
& & \partial_t (\gamma^{1/2} E) 
+ \partial_j (\gamma^{1/2} E v^j) 
= \nonumber \\
& & ~~~~~~ - P \left( \partial_t (\gamma^{1/2} W)
	+ \partial_i(\gamma^{1/2} W v^i) \right) \label{EWdot},
\end{eqnarray}
while the spatial components of (\ref{divT}) yield the Euler equation
\begin{eqnarray}
& & \partial_t (\gamma^{1/2} S_i) 
+ \partial_j (\gamma^{1/2} S_i v^j)
=  \nonumber \\
& & ~~~~~~ - \alpha \gamma^{1/2} \left( 
\partial_i P + \frac{S_a S_b}{2 \alpha S^t}\partial_i g^{ab} \right).
\label{SWdot}
\end{eqnarray}
For gamma-law equations of state
\begin{equation}
P = (\Gamma - 1) \rho_0 \epsilon,
\end{equation}
the right hand side of the energy equation (\ref{EWdot}) can 
be eliminated to yield
\begin{equation}  \label{ESdot}
\partial_t (\gamma^{1/2} E_*) 
+ \partial_j (\gamma^{1/2} E_* v^j) 
= 0,
\end{equation}
where we have introduced the energy variable $E_*$ defined as
\begin{equation}
E_* \equiv (\rho_0 \epsilon)^{1/\Gamma} W
\end{equation}
(see Shibata 1999, who also absorbed the determinant $\gamma^{1/2}$
into the definition of the fluid variables; see also Shibata, Baumgarte \&
Shapiro 1998).  This simplification has great computational
advantages, since the time derivatives on the right hand side of
(\ref{EWdot}) are difficult to handle in strongly relativistic fluid
flow (compare Norman \& Winkler 1986).

For given values of $u_i$, $W$ can be found from the normalization
relation $u_a u^a = -1$,
\begin{equation}
W = \alpha u^t = \left(1 + \gamma^{ij} u_i u_j \right)^{1/2},
\end{equation}
and $v^i$ from
\begin{equation}
v^i = \alpha \gamma^{ij}u_j / W - \beta^i.
\end{equation}

Finite difference implementations of the above equations must be
adapted to handle the appearance of shock discontinuities.  Two
strategies are commonly adopted.  

The more traditional approach is to add an artificial viscosity term
to the equations (von Neuman \& Richtmyer 1950).  Typically, the
artificial viscosity term $Q$ is non-zero only where the fluid is
compressed and is added to the pressure on the right hand sides of
both the energy equation (\ref{EWdot}) and the Euler equation
(\ref{SWdot}).  The artificial viscosity spreads the shock
discontinuity over several grid zones. For shocks occurring in
Newtonian fluids with modest Mach numbers, artificial viscosity
generates the Rankine-Hugoniot jump conditions to reasonable accuracy.
Artificial viscosity has also been used successfully in relativistic
applications (e.g.~Wilson 1972; Shibata 1999), but it leads to less
satisfactory results for highly relativistic flows or high Mach
numbers (Norman \& Winkler 1986).

An alternative approach to handling shocks is a high resolution shock
capturing (HRSC) scheme (see, e.g., Mart\'{\i} \& M\"uller 1999 for a
recent review).  In such a scheme, one treats all fluid variables as
constant in each grid cell. The discontinuous fluid variables at the
grid interfaces serve as initial conditions for a local Riemann shock
tube problem, which can be solved either exactly or
approximately. Allowing for discontinuities, including shocks, lies at
the core of these schemes, and does not require any additional
artificial viscosity.  Constructing Riemann solvers for HRSC requires
knowledge of the local characteristic structure of the equations to be
solved.  This has motived the development of several flux-conservative
hydrodynamics schemes, which do not contain any derivatives of the
fluid variables in the source terms, and for which this characteristic
structure can be determined (see, e.g., Font 2000; Font et al.~2002).

\section{General Relativistic Magnetohydrodynamics}
\label{sec6}

To derive the equations of general relativistic magnetohydrodynamics,
we now add the electromagnetic stress-energy tensor $T^{ab}_{\rm em}$
to the fluid stress-energy tensor
\begin{equation}
T_{ab} = T^{ab}_{\rm fluid} + T^{ab}_{\rm em}.
\end{equation}
Local conservation of energy-momentum demands that the divergence of the
sum $T_{ab}$ vanish. The divergence of the
individual stress-energy tensors $T^{ab}_{\rm fluid}$ and $T^{ab}_{\rm
em}$ do not vanish in general, since the fluid and electromagnetic 
fields may
exchange energy and momentum.  In particular, one finds
\begin{equation} \label{lorentz1}
\nabla_b T^{ab}_{\rm fluid} = - \nabla_b T^{ab}_{\rm em} =  F^{ab} {\mathcal J}_b
\end{equation}
(see equation (5.40) in Misner, Thorne \& Wheeler 1973).  The right
hand side of equation (\ref{lorentz1}) now includes the Lorentz force
in the equations of relativistic hydrodynamics. Note that the baryon
conservation equations (\ref{cont}) and (\ref{DWdot}) remain
unchanged.

In the energy equation (\ref{EWdot}), which was derived from the
contraction $u_b \nabla_a T^{ab}$, the addition of the Lorentz force
yields
\begin{eqnarray}
& & \partial_t (\gamma^{1/2} E) 
+ \partial_j (\gamma^{1/2} E v^j) 
=  \label{EWdot2} \\
& & ~~ - P \left( \partial_t (\gamma^{1/2} W)
	+ \partial_i(\gamma^{1/2} W v^i) \right) 
	- \alpha \gamma^{1/2} u_a F^{ab} {\mathcal J}_b.
	\nonumber
\end{eqnarray}
For a gamma-law equation of state the above equation may be written 
in terms of $E_*$ according to
\begin{eqnarray} 
& & \partial_t (\gamma^{1/2} E_*) 
+ \partial_j (\gamma^{1/2} E_* v^j) 
=  \label{ESdot2} \\
& & ~~ - {u_a F^{ab}  {\mathcal J}_b} 
\left( \frac{E_*}{W} \right)^{1 - \Gamma}
\left( \frac {\alpha \gamma^{1/2}}{\Gamma} \right),
\nonumber
\end{eqnarray}
which now takes the place of (\ref{ESdot}) in general.  The Euler
equation (\ref{SWdot}) now becomes
\begin{eqnarray}
& & \partial_t (\gamma^{1/2} S_i) 
+ \partial_j (\gamma^{1/2} S_i v^j)
= \label{SWdot2}  \\
& & ~~ - \alpha \gamma^{1/2} \left( 
\partial_i P + \frac{S_a S_b}{2 \alpha S^t}\partial_i g^{ab} \right)
+ \alpha \gamma^{1/2} F_{ia} {\mathcal J}^a. \nonumber
\end{eqnarray}

In the case of ideal MHD the new terms on the right hand sides of the
energy equations (\ref{EWdot2}) and (\ref{ESdot2}) vanish because of
(\ref{ideal_MHD}).  This result is understandable, since it
corresponds to the absence of Joule heating in the limit of infinite
conductivity.

We now proceed to determine the Lorentz force $F_{ia} {\mathcal J}^a$
in the Euler equation (\ref{SWdot2}).  Since ${\mathcal J}^a$ is not
known a priori, we will first use (\ref{maxwell1}) to express
${\mathcal J}_a$ in terms of the electromagnetic fields.  Note that
${\mathcal J}_a$ is the current four-vector as opposed to its spatial
projection $J^i$. We could express $J^i$ immediately using the spatial
Maxwell equation (\ref{Edot}).  Instead we need to derive the
four-dimensional equivalent of (\ref{Edot}) to express ${\mathcal
J}_a$, which we then can contract with the Faraday tensor
(\ref{faraday1}) to obtain the Lorentz force.

The divergence of the Faraday tensor in (\ref{lorentz1}) is
\begin{eqnarray} \label{divF}
\nabla_b F^{ab} & = & n^a \nabla_b E^b + E^b \nabla_b n^a
	- n^b \nabla_b E^a - E^a \nabla_b n^b \nonumber \\
& & + \nabla_b \epsilon^{abc} B_c.
\end{eqnarray}
We will now decompose the four-dimensional derivatives $\nabla_a$ in
each term above.

The Lie derivative of $E^a$ along $\alpha n^a$ is
\begin{equation}
\Lie_{\alpha {\bf n}} E^a 
= \alpha n^b \nabla_b E^a - \alpha E^b \nabla_b n^a - E^b n^a \nabla_b \alpha,
\end{equation}
or, with (\ref{t}),
\begin{equation}
\frac{1}{\alpha} (\partial_t - \Lie_{\bf \beta}) E^a
= n^b \nabla_b E^a - E^b \nabla_b n^a - E^b n^a \nabla_b \ln \alpha.
\end{equation}
The four-dimensional divergence $\nabla_a E^a$ can be expressed in
terms of the three-dimensional divergence $D_i E^i$
\begin{eqnarray}
D_a E^a & = & \gamma_a^{~b} \nabla_b E^a =
	(g_a^{~b} + n_a n^b) \nabla_b E^a \nonumber \\
	& = & \nabla_a E^a - E^a D_a \ln \alpha,
\end{eqnarray}
where we have used $n_a E^a = 0$ and
\begin{equation} \label{a}
n^b \nabla_b n_a = a_a = D_a \ln \alpha.
\end{equation}
Here $a_a$ is the four-acceleration of a normal observer.
Since the extrinsic curvature $K_{ab}$ can be written
\begin{equation} \label{K}
K_{ab} = - \nabla_a n_b - n_a a_b,
\end{equation}
the divergence of $n^a$ satisfies
\begin{equation}
\nabla_a n^a = - K.
\end{equation}
Inserting these expressions into (\ref{divF}), we now find the
intermediate result
\begin{eqnarray} \label{divF2}
\nabla_b F^{ab} & = & n^a D_b E^b + K E^a 
	- \frac{1}{\alpha}(\partial_t - \Lie_{\bf \beta}) E^a \nonumber \\ 
& & + \nabla_b \epsilon^{abc} B_c,
\end{eqnarray}
where we have also used $ E^a D_a \ln \alpha = E^a \nabla_a \ln
\alpha$.

The term involving $B_a$ in (\ref{divF2}) can be rewritten as
\begin{eqnarray}
\nabla_b \epsilon^{abc} B_c & = &
\epsilon^{abcd} \nabla_b B_c n_d
= \epsilon^{abcd} (n_d \nabla_b B_c + B_c \nabla_b n_d ) \nonumber \\
& = & \epsilon^{abcd} (n_d D_b B_c - B_c n_b a_d) \nonumber \\
& = & \alpha^{-1} \epsilon^{abcd} (\alpha n_d D_b B_c + n_d B_c D_b \alpha) 
\nonumber \\
& = & \alpha^{-1} \epsilon^{abc} D_b (\alpha B_c),
\end{eqnarray} 
where we have used 
equations (\ref{epsilon}), (\ref{a}) and (\ref{K}).  Inserting this 
expression into (\ref{divF2})
now yields
\begin{eqnarray} \label{J}
4 \pi \alpha {\mathcal J}^a = \alpha \nabla_b F^{ab} & = &   
	- (\partial_t - \Lie_{\bf \beta}) E^a 
	+ \epsilon^{abc} D_b (\alpha B_c) \nonumber \\
	& & \alpha  n^a D_b E^b
	+  \alpha K E^a.
\end{eqnarray}
Not surprisingly, this is the four-dimensional version of equation
(\ref{Edot}), which can be found by taking the spatial projection of
(\ref{J}).

The next step is to contract (\ref{J}) with the Faraday tensor
(\ref{faraday1}).  Using $n_a E^a=0$, $n_a \epsilon^{abc} = 0$ 
and $n_a \Lie_{\alpha {\bf n}} E^a = 0$ several terms cancel, and one
finds
\begin{eqnarray}
& & 4 \pi \alpha {\mathcal J}^b F_{ab} = \alpha E_a D_b E^b \label{J2} \\
& & ~~+ n_a \left( - E_b (\partial_t - \Lie_{\bf \beta} - \alpha K) E^b   
	+ E_b \epsilon^{bcd} D_c (\alpha B_d) \right) \nonumber \\
& & ~~- B^c \epsilon_{cab}
	\left( (\partial_t - \Lie_{\bf \beta} - \alpha K) E^b
	- \epsilon^{bde} D_d (\alpha B_e) \right). \nonumber
\end{eqnarray}
This expression can now be inserted into the Euler equation
(\ref{SWdot2}).  For spatial components $n_i = 0$, so that the second
line in (\ref{J2}) vanishes.  The source term $\alpha \gamma^{1/2}
F_{ia} {\mathcal J}^a$ can then be rewritten
\begin{eqnarray}
& & \alpha \gamma^{1/2} F_{ia} {\mathcal J}^a =
\frac{\gamma^{1/2}}{4 \pi} \Big( - B^j \epsilon_{jik} 
(\partial_t - \Lie_{\bf \beta} - \alpha K) E^k \nonumber \\ 
& & ~~ + \alpha E_i (D_j E^j) 
+ B^j D_j(\alpha B_i) - B^j D_i (\alpha B_j) \Big). 
\end{eqnarray}
If desired, the covariant derivatives in the last two terms can be
converted into partial derivatives, which finally yields the Euler
equation
\begin{eqnarray} 
& & \partial_t (\gamma^{1/2} S_i) 
+ \partial_j (\gamma^{1/2} S_i v^j)
= \label{SWdot3} \nonumber \\
& & ~~ - \alpha \gamma^{1/2} \left( 
\partial_i P + \frac{S_a S_b}{2 \alpha S^t}\partial_i g^{ab} \right)
+ \alpha \frac{\gamma^{1/2}}{4 \pi}  E_i (D_j E^j) \nonumber \\
& & ~~ - \frac{\gamma^{1/2}}{4 \pi} B^j
\Big(\epsilon_{jik} (\partial_t E^k  - \beta^l \partial_l E^k
+ E^l \partial_l \beta^k - \alpha K E^k) \nonumber \\
& & ~~ + \partial_i (\alpha B_j) -  \partial_j(\alpha B_i) \Big),
\end{eqnarray}
where we have expanded the Lie derivative of $E^i$.  Note that in this
equation the electric field terms enter with the opposite sign from
those in the corresponding equation (3.2) of Sloan \& Smarr (1985, hereafter
SS), who further assume $\beta = 0 = K$.

For numerical implementations, the most challenging term in Euler's
equation is probably the time-derivative of the electric field.  For
ideal MHD, this term can be rewritten by first expressing $E^i$ in
terms of the magnetic fields $B^i$ using the ideal MHD relation
(\ref{MHD}), and then using (\ref{MHD_faraday}) to eliminate the
time-derivative of $B^i$ (see Zhang 1989). This term is likely to be
small in most applications; for example, it is $\mathcal{O}(v^2/c^2)$
times smaller than the last two terms on the right hand side of
(\ref{SWdot3}).  In such cases, extrapolating and iterating, or some 
other simple treatment, may be adequate to account for its contribution.

It is instructive to take the Newtonian limit of equation
(\ref{SWdot3}) and recover a familiar expression.  With $g_{00}
\rightarrow -(1 + 2 \phi)$, where $\phi$ is the Newtonian potential,
we find
\begin{equation}
\frac{1}{2} \frac{S_a S_b}{\alpha S^t} \partial_i g^{ab} 
\rightarrow - \frac{1}{2} \rho \partial_i g^{00}
= \rho \partial_i \phi.
\end{equation}
In cartesian coordinates ($\gamma^{1/2} = 1$), the Newtonian limit of
equation (\ref{SWdot3}) then becomes
\begin{eqnarray}
& & \partial_t S_i 
+ \partial_j ( S_i v^j)
=  \label{SWdotNewton} 
- \partial_i P - \rho \partial_i \phi 
+ \rho_e E_i \nonumber \\
& &  ~~~ - \frac{1}{8 \pi} \partial_i (B^j B_j) 
+ \frac{1}{4 \pi} B^j \partial_j B_i
\end{eqnarray}
or equivalently
\begin{equation}
\rho \frac{d {\bf v}}{dt} =
- {\bf \nabla} (P + P_{\rm M}) - \rho {\bf \nabla} \phi
+ \frac{1}{4 \pi} ({\bf B} \cdot {\bf \nabla}) {\bf B} + \rho_e {\bf E},
\end{equation}
where we have defined the magnetic pressure
\begin{equation}
P_{\rm M} = \frac{{\bf B}^2}{8 \pi}
\end{equation}
and where ${\bf \nabla}$ is the spatial gradient operator.  Note that
for a neutral plasma $\rho_e = 0$ the electric field $E^a$ disappears
entirely from the above Newtonian equation.

\section{Source Terms for the Gravitational Field Equations}
\label{sec7}

We now catalogue the source terms $\rho$ (\ref{rho}), $S_i$
(\ref{Si}), $S_{ij}$ (\ref{Sij}) and $S$ (\ref{trS}) that appear in
the Hamiltonian constraint (\ref{ham}), the momentum constraint
(\ref{mom}) and the evolution equation (\ref{kdot}). Inserting the
fluid stress-energy tensor (\ref{T_fluid}) into equations (\ref{rho})
- (\ref{trS}) yields the fluid contributions to the source terms:
\begin{eqnarray}
\rho_{\rm fluid} & = & \rho_0 h W^2 - P \label{rhof} \\
S_i^{\rm fluid} & = & \rho_0 h W u_i \label{Sif} \\
S_{ij}^{\rm fluid} & =& P \gamma_{ij} + \frac{S_i^{\rm fluid}S_j^{\rm fluid}}
    {\rho_0 h W^2} \label{Sijf} \\
S_{\rm fluid} & = & 3P + \rho_0 h (W^2 - 1). \label{Sf}
\end{eqnarray}

Next we assemble the electromagnetic contributions to the source
terms.  To do so, we first need to construct the electromagnetic
stress-energy tensor $T^{ab}_{\rm em}$ from the Faraday tensor
$F^{ab}$,
\begin{equation} \label{Tem}
4 \pi T^{ab}_{\rm em} = F^{ac} F^b_{~c} - \frac{1}{4} g^{ab} F_{cd} F^{cd}.
\end{equation} 
Inserting (\ref{faraday1}), we first find 
\begin{equation}
F_{ab} F^{ab} = 2 (B_i B^i - E_i E^i),
\end{equation}
where we have used $\epsilon_{abc}\epsilon^{abd} = 2 \gamma_c^{~d}$.
With $\epsilon^{abc} \epsilon_a^{~de} = \gamma^{bd} \gamma^{ce} -
\gamma^{be} \gamma^{cd}$, the first term in (\ref{Tem}) becomes
\begin{eqnarray}
F^{ac} F^b_{~c} & = & n^a n^b E_i E^i + 2 n^{(a} \epsilon^{b)cd} E_c B_d
	\nonumber \\
& & - E^a E^b - B^a B^b + \gamma^{ab} B_i B^i.
\end{eqnarray}
Combining the last two equations then yields the electromagnetic
stress-energy tensor in $3+1$ form
\begin{eqnarray}
4 \pi T^{ab}_{\rm em} & = & 
\frac{1}{2} (n^a n^b + \gamma^{ab}) (E_i E^i + B_i B^i)
\nonumber \\
& & + 2 n^{(a} \epsilon^{b)cd} E_c B_d - (E^a E^b + B^a B^b).
\end{eqnarray}

This stress-energy tensor can now be inserted into (\ref{rho}) to
(\ref{trS}) to obtain the electromagnetic source terms.  For the
mass-energy density $\rho_{\rm em}$ we find
\begin{equation}
4 \pi \rho_{\rm em} = n_a n_b 4 \pi T^{ab}_{\rm em} = \frac{1}{2} (E_i E^i + B_i B^i)
= \frac{1}{2} ({\bf E}^2 + {\bf B}^2),
\end{equation}
which is the energy density of the electromagnetic fields.  The energy
flux $S_i^{\rm em}$ reduces to the Poynting vector
\begin{eqnarray}
4 \pi S_i^{\rm em} & = & - \gamma_{ia} n_b 4 \pi T^{ab}_{\rm em} = 
	- \gamma_{ia} n_b n^b \epsilon^{acd} E_c B_d \nonumber \\
& = & \epsilon_{ijk} E^j B^k = ({\bf E \times B})_i.
\end{eqnarray}
The stress tensor $S_{ij}^{\rm em}$ is
\begin{equation}
4 \pi S_{ij}^{\rm em} = \gamma_{ia} \gamma_{jb} 4 \pi T^{ab}_{\rm em}
	= - E_i E_j - B_i B_j + \frac{1}{2} \gamma_{ij} ({\bf E}^2 +
	{\bf B}^2).
\end{equation}
Its trace (\ref{trS}), finally, is equal to the mass-energy 
density $\rho_{\rm em}$
\begin{equation}
4 \pi S_{\rm em} = \frac{1}{2} ({\bf E}^2 + {\bf B}^2).
\end{equation}
The above results are not surprising: expressed in terms of the
electromagnetic field components as measured by a normal observer,
$n^a$, i.e.~an observer who is at rest with respect to the slices
$\Sigma$, the $3+1$ source terms have the same form as in flat space
(compare exercise 5.1 in Misner, Thorne \& Wheeler 1973).

\section{Comparison with Previous Treatments}
\label{sec8}

In this section we compare our notation and findings with those of
Sloan \& Smarr (1985, SS), Hawley \& Evans (1988, EH; and 1989, HE)
and Zhang (1989, Z).

SS define the three-velocity $v^i_{\rm SS}$ by writing the
four-velocity $u^a$ as
\begin{equation} \label{uSS}
u^a = \alpha^{-1} W (1, \alpha v^i_{\rm SS} - \beta^i)
\end{equation}
and
\begin{equation} \label{ulowSS}
u_a = W (- \alpha + v^{\rm SS}_i \beta^i,  v^{\rm SS}_i )
\end{equation}
(see equations (SS-2.1)).  Here $W$ is the Lorentz factor between
$u^a$ and $n^a$
\begin{equation}
W = - n_a u^a = \alpha u^t
\end{equation}
as in (\ref{W}). The normalization $u^a u_a = -1$ leads to
\begin{equation}
W = (1 - v^i_{\rm SS}v_i^{\rm SS})^{-1/2}
\end{equation}
which shows that $v^i_{\rm SS}$ is the velocity of the fluid with
respect to a {\it normal} observer.

Zhang (1989) adopts the same formulation as SS, but denotes $W$ with
$\gamma$ (see equation (Z-2.11)).

HE adopt the same definition of three-velocity as we do, defining the
three-velocity $v^i$ to be the velocity with respect to {\it
coordinate} observers,
\begin{equation} \label{vW}
v^i_{\rm W} = u^i/u^t,
\end{equation}
(see equation (\ref{v}) above).  We use the subscript W since this
definition is used in Wilson's equations of relativistic hydrodynamics
(see Wilson 1972; Hawley, Smarr \& Wilson 1984).  With (\ref{vW}), the
four-velocity $u^a$ can be written
\begin{equation} \label{uW}
u^a = \alpha^{-1} W(1, v^i_{\rm W}).
\end{equation}
Comparing (\ref{uSS}) and (\ref{uW}) shows that the two definitions
of $v^i$ are related by
\begin{equation} \label{translate_v}
v^i_{\rm W} = \alpha v^i_{\rm SS} - \beta^i.
\end{equation}

We can now compare the ideal MHD equation (\ref{MHD}) in the different
treatments.  Since HE adopt the same definition for $v^i = v^i_{\rm
W}$ as we do, their equations (EH-A14) and (HE-14) should be identical
to our equation (\ref{MHD}).  In their expression, however, the shift
term is absent.  This absent shift term can be traced back to their
equation (HE-13), which does not agree with our equation
(\ref{ohm_spatial}).  It is likely that the shift term was missed by
dropping the term $\epsilon_{itj}$. The alignment of indices is
incorrect in SS's equation (2.9) (which they express in terms of $u^i$
instead of $v^i$).  Fixing it and utilizing (\ref{ulowSS}) and
(\ref{translate_v}) makes their equation equivalent to (\ref{MHD}).

To compare with Zhang's ideal MHD equation (Z-2.12), we insert
(\ref{translate_v}) into (\ref{MHD}), which immediately yields
Zhang's result
\begin{equation}
E_i = - \epsilon_{ijk} v_{\rm SS}^j B^k,
\end{equation}
showing that our result agrees with that of Zhang.

We find similar errors in the Faraday equation.  We found that the
shift terms in (\ref{MHD}) cancel all other shift terms when inserted
into (\ref{Bdot}), ultimately yielding expressions (\ref{MHD_faraday})
and (\ref{MHD_faraday2}) which do not include any shift terms.  With
the shift terms being absent in equation (HE-14), the corresponding
terms do not cancel, leading to the incorrect equations (HE-17) and
(HE-18) (see also (EH-2.8) and (EH-A17)).

Zhang's expression for the Faraday equation (Z-2.13') can be recovered
by inserting (\ref{translate_v}) into (\ref{MHD_faraday2})
\begin{equation} 
\partial_t {\mathcal B}^i = 
	\partial_j \left( (\alpha v_{\rm SS}^i - \beta^i) {\mathcal B}^j - 
	(\alpha v_{\rm SS}^j - \beta^j) {\mathcal B}^i \right),
\end{equation}
which can be rewritten as
\begin{equation}
\frac{1}{\gamma^{1/2}} \partial_t (\gamma^{1/2} B^i) =
D_j  \left( (\alpha v_{\rm SS}^i - \beta^i) {B}^j - 
	(\alpha v_{\rm SS}^j - \beta^j) {B}^i \right)
\end{equation}
or
\begin{equation}
\frac{1}{\gamma^{1/2}} \partial_t (\gamma^{1/2} {\bf B}) =
\nabla \times \left( (\alpha {\bf v}_{\rm SS} - {\bf \beta}) \times {\bf B} 
	\right).
\end{equation}
This shows that our equations (\ref{MHD_faraday}) and
(\ref{MHD_faraday2}) again agree with the expressions of Zhang.

Interestingly, Zhang refers to EH, and in fact their equations look
quite similar in that they both contain the above shift terms.
However, Zhang uses $v^i_{\rm SS}$ as the three-velocity, while Hawley
and Evans use $v^i_{\rm W}$.  Therefore the shift terms are correct in
the former, but incorrect in the latter.

Finally, we show that our equation (\ref{ohm_spatial}) is equivalent
to Zhang's equation (Z-2.10).  On the left hand side of
(\ref{ohm_spatial}) we rewrite
\begin{eqnarray}
\tilde \rho_e & = & u_a {\mathcal J}^a = u_a ( n^a \rho_e + J^a )
= - W \rho_e + u_i J^i \nonumber \\
& = &  W ( {\bf v}_{\rm SS} \cdot {\bf J} - \rho_e)
\end{eqnarray}
where we have used the decomposition (\ref{defJ}).  Inserting this 
and (\ref{translate_v}) into (\ref{ohm_spatial}) yields
\begin{equation}
{\bf J} + W^2 ({\bf v}_{\rm SS} \cdot {\bf J} - \rho_e) {\bf v}_{\rm SS}
= \sigma W ( {\bf E} + {\bf v}_{\rm SS} \times {\bf B}),
\end{equation}
which is identical to (Z-2.10).

\section{Summary}
\label{sec9}

We have assembled a complete set of Maxwell-Einstein-MHD equations,
describing the structure and evolution of a relativistic, ideal MHD
gas in a dynamical spacetime.  We compare with previous treatments,
and correct some errors in the existing literature.

Our compilation of these equations is motivated by a large number of
problems in relativistic astrophysics in which magnetic fields are
likely to play an important role (see the incomplete list in Section
\ref{sec1}).  Self-consistent solutions to the Maxwell-Einstein-MHD
equations will be necessary for a thorough understanding of these
problems, and we therefore anticipate that relativistic MHD in
dynamical spacetimes will attract much interest in the future.  We
hope that our compilation of these equations will be useful for such
investigations, particularly for treatments that will rely on
numerical simulations.  In Paper II, we use these equations to model
the collapse of a magnetized star to a black hole.

\acknowledgments

It is a pleasure to thank C. Gammie for useful discussions.  This work
was supported in part by NSF Grants PHY-0090310 and PHY-0205155 and
NASA Grant NAG 5-10781 at the University of Illinois at
Urbana-Champaign and NSF Grant PHY 0139907 at Bowdoin College.

\end{document}